\documentclass[%
 reprint,
 amsmath,amssymb,floatfix,
 aps,
]{revtex4-1}

\usepackage{graphicx}
\usepackage{dcolumn}
\usepackage{bm}
\usepackage{comment}
\usepackage{mathtools}

\begin{document}

\preprint{APS/123-QED}

\title{A Generic Rotating-Frame-Based Approach to Chaos Generation in Nonlinear M/NEMS Resonators}

\author{Samer Houri}
\email{Samer.Houri.dg@hco.ntt.co.jp}
  \affiliation{NTT Basic Research Laboratories, NTT Corporation,
3-1 Morinosato-Wakamiya, Atsugi-shi, Kanagawa 243-0198, Japan.}

\author{Motoki Asano}
\author{Hiroshi Yamaguchi}
  \affiliation{NTT Basic Research Laboratories, NTT Corporation,
3-1 Morinosato-Wakamiya, Atsugi-shi, Kanagawa 243-0198, Japan.}
\author{Natsue Yoshimura}%
\affiliation{%
Institute of Innovative Research, Tokyo Institute of Technology, Yokohama 226-8503, Japan.
}%
\author{Yasuharu Koike}%
\affiliation{%
Institute of Innovative Research, Tokyo Institute of Technology, Yokohama 226-8503, Japan.
}%

\author{Ludovico Minati}%
 \email{minati.l.aa@m.titech.ac.jp}
\affiliation{%
Tokyo Tech World Research Hub Initiative (WRHI), Tokyo Institute of Technology, Yokohama 226-8503, Japan.
}%

\date{\today}

\begin{abstract}

This work provides a low-power method for chaos generation which is generally applicable to nonlinear M/NEMS resonators. The approach taken is independent of the material, scale, design, and actuation of the device in question; it simply assumes a good quality factor and a Duffing type nonlinearity, features that are commonplace to M/NEMS resonators. The approach models the rotating-frame dynamics to analytically constrain the parameter space required for chaos generation. By leveraging these common properties of M/NEMS devices, a period-doubling route to chaos is generated using an order-of-magnitude smaller forcing than typically reported in the literature.\\
\end{abstract}

\maketitle

\indent\indent Chaotic dynamics have received interest 
owing to their extraordinary ability to generate complex behaviors, such as synchronization patterns, even in simple and fixed arrangements of coupled nodes. Countless applications have been discussed, spanning control systems, telecommunications and neuroscience \cite{hilborn2000chaos,ott2002chaos,buscarino2014concise,boccaletti2018synchronization}. Recently, the field has witnessed a resurgence of interest due to the possibility of building large-scale hardware reservoirs from coupled nonlinear oscillators. To meet the requirements for practical application, high-integration and low-power implementation are necessary \cite{minati2018across,tanaka2019recent}.\\
\indent\indent\ Micro- and nano-electromechanical systems (M/NEMS) provide experimental platforms for investigating and generating such dynamics, as they are easily amenable to very large-scale integration, low-power operation, and they inherently exhibit rich nonlinear behavior \cite{lifshitz2008nonlinear}. Indeed, chaos generation is well-reported in the M/NEMS scientific literature \cite{guttinger2017energy,karabalin2009nonlinear,kenig2011homoclinic,bienstman1998autonomous,ashhab1999melnikov,park2008energy,de2006complex,wang1998chaos,demartini2007chaos,towfighian2011analysis}. However, in the existing studies, chaos generation was obtained in a manner that is system/device specific and neither provides a minimalist approach nor a general one to generating chaos in such devices. What is meant here by general is a procedure that is device-independent, i.e., an approach that would reliably generate chaos without detailed knowledge of the material, actuation, dimensions, or design of M/NEMS devices. The term minimalist 
indicates the smallest number of phase-space dimensions ($n$) necessary for the onset of chaos (i.e., $n=3$).\\
\indent\indent\ Chaos generation reported in the literature for M/NEMS devices resorts to non-general properties, e.g. non-smooth nonlinearity \cite{bienstman1998autonomous}, a high number of phase-space dimensions ($n>3$) obtained via inter-device or inter-modal coupling \cite{karabalin2009nonlinear,guttinger2017energy,kenig2011homoclinic}, or extreme nonlinearity by 
operating near the electrostatic pull-in \cite{park2008energy,de2006complex}. However, a 
common approach is based on the creation of a static double-well potential either through electrostatic forces or by using buckled structures \cite{wang1998chaos,demartini2007chaos,towfighian2011analysis,emam2004nonlinear}. None of these approaches hinges around a common denominator property: they are, therefore, not transferable to nonlinear M/NEMS resonators in general. In fact, they commonly require 
a large actuation voltage, negating one of the main appeals of M/NEMS devices.\\
\indent\indent\ An 
interesting case, is that of the static double-well potential, which is 
described by a Duffing equation (
a system possessing a cubic nonlinearity)
where the linear component is negative and the cubic component is positive. Period-doubling bifurcations and chaos in such systems have been 
studied \cite{sharma2012effects} and been subject to 
experimental investigations \cite{moon1980experiments,holmes1979nonlinear,moon1979magnetoelastic}. While such systems can be reproduced in M/NEMS devices \cite{wang1998chaos,demartini2007chaos,towfighian2011analysis,emam2004nonlinear}, nearly all M/NEMS devices inherently exhibit a different type of cubic nonlinearity \cite{kaajakari2004nonlinear}, equally captured by a variant of the Duffing equation, whereby the linear component is positive and the cubic component can be either positive, or negative. Such 
M/NEMS resonators can 
exhibit dynamic bistability \cite{cleland2013foundations}.\\
\indent\indent\ In contrast to the static double-well systems, the approach presented here relies on the dynamically-generated double-well pseudo-potential that is created when a generic nonlinear Duffing resonator is driven into the bistable regime \cite{cleland2013foundations}. Since bistability is accessed when the resonator's vibration amplitude is on the order of some scaling parameter (e.g. thickness \cite{kaajakari2004nonlinear}), such pseudo-potentials can be generated and manipulated by 
changing the drive conditions without requiring device or setup redesign.\\
\indent\indent\ The main purpose of this work is to demonstrate 
chaos generation in a perturbed ``dynamic double-well" in a manner that parallels chaos generation in driven ``static double-well" systems, albeit with significantly lower drive amplitudes. Thus, two drive tones are applied to the system, whereby the first creates the ``dynamic double-well" and the second perturbs it.\\
\indent\indent\ Since the displacement of the M/NEMS resonator studied herein is moderate, i.e. on the order of the structural thickness, a perturbation-based approach to analyse the dynamics is justified \cite{cleland2013foundations,greywall1994evading,kozinsky2007basins}. To represent the underlying dynamics we employ the rotating frame approximation (RFA), whereby a ``slow flow” (a time-varying envelope) dynamics is overlaid on top of an otherwise purely sinusoidal response, and the timescales ($\mathcal{\tau}$) associated with this “slow flow” are on the order of the resonator line-width ($\mathcal{\gamma}$), i.e., $\mathnormal{\tau\approx}~\mathcal{O}\mathnormal{(1/\gamma)}$ \cite{greywall1994evading,cleland2013foundations}.\\
\indent\indent\ When a Duffing resonator is driven near resonance, its steady-state response as seen in the RFA corresponds to a fixed-point in the phase space; in case the resonator is driven in the bistable regime, the response shows two distinct stable fixed points and a saddle-node point \cite{kozinsky2007basins}. This latter configuration implies 
a homoclinic connection (i.e., a trajectory that starts and ends in the saddle-node and orbits one of the stable fixed points) may exist in the RFA phase-space of a 
Duffing resonator. Thus, just as a ``static double-well" potential provides a homoclinic connection in the rest frame, so does the dynamic Duffing bistability provide a homoclinic connection in the rotating-frame, and it is the perturbation of such homoclinics that is responsible for the generation of chaos \cite{mel1963stability, holmes1979nonlinear,sharma2012effects}.\\
\indent\indent\ This argument can be demonstrated by considering the Duffing equation for a two tone-driven 
M/NEMS resonator, given as 
\cite{cleland2013foundations}\\
\begin{multline}
{\ddot{x} + \gamma\dot{x} + \omega_{0}^2x + \alpha x^3 = {\eta F_1}\cos(\omega_1t) + {\eta F_2}\cos(\omega_2t)}\\
\label{eqn:Eq1}
\end{multline}
\indent\indent\ where $x$ is the displacement, and $\gamma$, $\omega_0$, $\alpha$ are respectively the damping, natural frequency, and Duffing nonlinearity of the resonator. ${F_1}$, ${F_2}$, $\omega_1$, and $\omega_2$ are the amplitudes and frequencies of two-externally applied driving forces, and $\eta$ is the transduction coefficient.
We introduce dimensionless constants 
as ${\bar{t}=t\times\omega_0}$, ${\bar{\gamma}=\gamma/\omega_0}$, ${\bar{\alpha}=\alpha/\omega_0^2}$, ${\bar{{F_1}}={\eta F_1}/\omega_0^2}$, and ${\bar{{F_2}}={\eta F_2}/\omega_0^2}$. Hereon, all equations are written using this form, however, the bars are dropped for convenience.\\
\indent\indent\ The application of the RFA, in which the modal displacement takes the form ${x(t)=A(t)\cos(\omega_1 t+\phi(t))}$, where ${A(t)}$ and ${\phi(t)}$ are slowly varying amplitude and phase envelopes, gives the following rotating-frame system\\
\begin{eqnarray}
\begin{cases}
{\dot{X}=\delta Y - \frac{3}{8}\alpha A^{2}Y + \frac{1}{2}({F_2}\sin(\Theta) - \gamma X)}\\
{\dot{Y}=-\delta X + \frac{3}{8}\alpha A^{2}X - \frac{1}{2}F_1 -  \frac{1}{2}({F_2}\cos(\Theta) + \gamma Y)}\\
{\dot\Theta=\Omega=(\omega_2-\omega_1)/\omega_0}\\
\end{cases}
\end{eqnarray}

\indent\indent\ where ${X = A\cos(\phi)}$ and ${Y = A\sin(\phi)}$ are the rotating-frame quadratures, ${A^2 = X^2 + Y^2}$, and ${\delta=(\omega_1 -\omega_0})/\omega_0$ (
details are provided in the supplementary materials).\\
\indent\indent\ Since Eqs.~(1)-(2) are generically applicable to Duffing-type resonators, the results below can, in principle, be implemented in various physical realizations of nonlinear resonators, such as optical \cite{hansson2014bichromatically} and superconducting resonators \cite{manucharyan2007microwave}, without loss of generality.\\
\indent\indent\ Initially, consider the conventional case with only one applied force, i.e., ${{F_2} = 0}$, whereby the system in Eq.~(2) is reduced to the first two equations only. The fixed points of the system, obtained by setting the time-derivative in Eq.~(2) to zero, exhibit a bistable response in a region of the dimensionless parameter space shown in Fig.~1(a) for a lossless and a low-loss ($\gamma=10^{-3}$) driven Duffing resonators. To visualize the phase space and associated homoclinic orbit, we select a constant-force cut through the parameter space, Fig.~1(b), and then a constant-detuning cut where bisatbility exists, Fig.~1(c). It is convenient to plot the RFA Hamiltonian ($\mathcal{H}_{\mathrm{RFA}}$) \cite{dykman2019resonantly,huber2019detecting}, shown in Fig.~1(c), along with the fixed points. For the case $\gamma=0$, trajectories on the $\mathcal{H}_{\mathrm{RFA}}$ surface follow closed orbits around the fixed points, the so-called libration orbits \cite{Houri2020}. A homoclinic orbit is then the limit case in which the libration orbit intersects the saddle-node point, as shown in Fig.~1(c).\\

\begin{figure}
	\graphicspath{{Figures/}}
	\includegraphics[width=85mm]{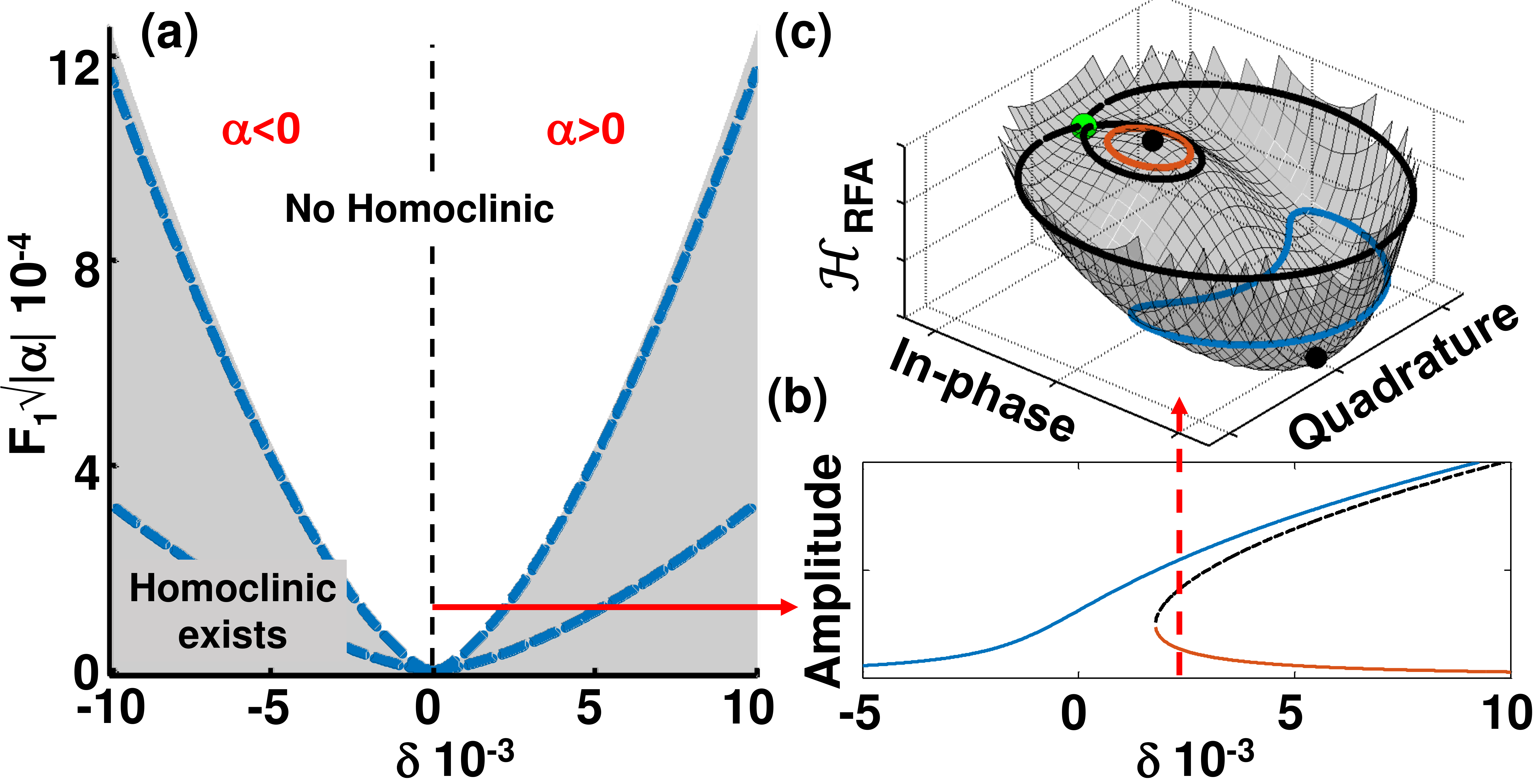}
	\caption{(a) Bistability map plotted as a function of dimensionless force and detuning, showing the region of bistability for a lossless driven Duffing resonator (grey area), and for a low-loss (Q=1000) Duffing resonator (area between the dashed blue lines). (b) Amplitude (arbitrary units) versus detuning response of a lossless Duffing taken for ${F_1\sqrt{\alpha} = 10^{-4}}$. The corresponding phase-space plot (also in arbitrary units) for a detuning of ${{\delta} = 2.5\times10^{-3}}$ is shown in (c). The stable fixed points and the saddle point are shown as black and green dots respectively, and the black traces correspond to the homoclinic orbit. Small amplitude libration orbits around the high-amplitude branch (blue) and low-amplitude branch (red) are shown. $\alpha$ is the Duffing parameter.}
\end{figure}

\indent\indent\ Note that for $F_2=0$ the system of Eq.~(2) is reduced to an autonomous two-dimensional system ($n=2$), which can not generate chaos as it lacks the necessary dimensionality. However, an additional time dependence introduced by making ${{F_2}\neq0}$ increases the RFA dimensions from $n=2$ to $n=3$, thus in principle meeting the condition for chaos generation. Thus, by setting ${{F_2}\neq0}$ chaos could be generated if the homoclinic orbit is sufficiently perturbed.\\
\indent\indent\ The above arguments are corroborated by a combination of numerical simulations and measurements. The experiments are performed using a micro-beam GaAs piezoelectric MEMS resonator driven into the nonlinear Duffing regime, details regarding fabrication and basic properties of such resonators can be found in Ref. \cite{yamaguchi2017gaas}. The resonator is placed in a vacuum chamber, and its motion is measured using a laser Doppler vibrometer, whose output is fed simultaneously to a lock-in amplifier and a vector signal analyzer (for experimental details, see supplementary information). A higher harmonic mode is selected for these experiments to avoid possible inter-modal interactions \cite{houri2019multimode,houri2019limit}.\\
\indent\indent\ The application of a single-tone sweep produces the frequency response shown in Fig.~2(a) for the linear (${100~\mathrm{mV_{PP}}}$, black trace) and nonlinear Duffing regimes (${3~\mathrm{V_{PP}}}$, red/blue traces), which, upon fitting, give the following values for the resonator parameters ${\omega_0/2\pi=1.56}$ MHz, $Q~=~{1000}$, and ${\alpha=1.67\times10^{15} \mathrm{Hz/V^2}}$.\\
\indent\indent As a first demonstration of the period-doubling route to chaos, a two-tone excitation is applied to the resonator with one fixed tone in the bistability region ($F_1~=~3~\mathrm{V_{PP}}$, $\omega_1/2\pi~=~1566.5$ kHz) and one swept tone. For large detunings between the two tones, the rotating-frame response corresponds to a low-amplitude libration oscillation having a frequency ${\Omega}$. As the tone is swept, the oscillations exhibit a quick succession of period-doubling bifurcations leading to chaos, as shown in Fig.~2(b) for both the high- and low- amplitude branches. The corresponding phase-space plots for period 1 (P1), period 2 (P2), and chaotic attractors are shown in Figs.~2(c)-(e), respectively. Note that, as will be demonstrated later, Fig.~2(b) indicates that the low-amplitude branch only generates chaos for negative detuning, i.e. $\Omega<0$, while the high-amplitude branch only generates chaos for $\Omega>0$.\\

\begin{figure}
	\graphicspath{{Figures/}}
	\includegraphics[width=85mm]{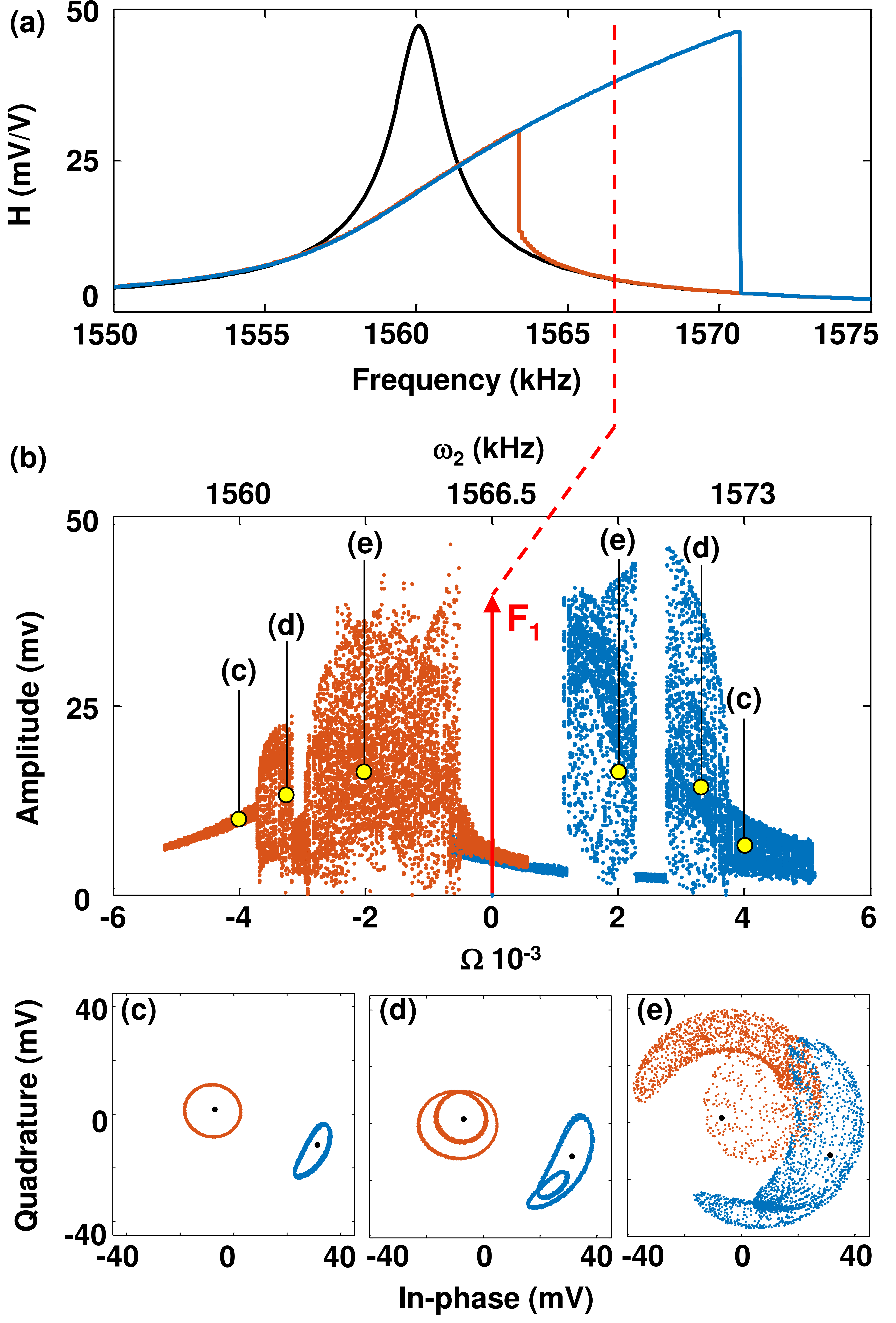}
	\caption{(a) Experimental lock-in amplifier data for the frequency response obtained by a single tone sweep showing the linear (black trace ${100~\mathrm{mV_{PP}}}$) and the Duffing regimes (${3~\mathrm{V_{PP}}}$), this latter shows bistability upon performing a forward (blue trace) and a backward sweep (red trace). \emph{H} denotes the relative amplitude response, expressed in mV per V drive. (b) Scatter plot of periodically-sampled amplitude of the rotating-frame oscillations under the effect of a two-tone excitation, with one fixed tone (indicated by ${F_1}$ having $\omega_1/2\pi~=~1566.5$ kHz and $F_1~=~3~\mathrm{V_{PP}}$) and one swept tone (${F_2}~=~2.1~\mathrm{V_{PP}}$, $1558$ kHz $<~\omega_2/2\pi~<~1576$ kHz), shown for the lower (red) and higher (blue) amplitude branches. Period 1, Period 2, and chaotic oscillations are detected and shown in (c)-(e) respectively, for both the high (blue) and low (red) amplitude branches. The black dots correspond to the experimentally obtained fixed points.}
\end{figure}

\indent\indent\ Making ${{F_2}\neq0}$ is clearly a necessary but not sufficient condition for chaos generation, and the question of whether and where in the 4-dimensional parameter space (${\delta}$, ${\Omega}$, ${F_1}$, and ${F_2}$) chaos exists remains to be addressed. While bounds have been set on the values of $F_1$ and $\delta$ such that bistability exists for $F_2=0$, similar bounds for $F_2$ and $\Omega$ have yet to be determined. This task is usually performed by applying Melnikov's method, which sets strict conditions for the period-doubling bifurcation route to chaos to take place \cite{moon1979magnetoelastic,sharma2012effects,holmes1979nonlinear,moon1980experiments,moon1984fractal,moon1985fractal}.\\
\indent\indent\ Ideally, the application of Melnikov's method constrains chaos generation in parameter-space with an analytical bound. Unfortunately, straightforward application of Melnikov's integral to the archetypal nonlinear resonator captured by Eq.~(1) results in a relation that is not easily amenable to analytical solution; therefore, a heuristic approach analogous to the one employed in Ref. \cite{moon1980experiments} is used. This approach considers that the application of a second forcing term creates the libration orbits, and the amplitude at which these libration orbits are large enough to undergo inter-well jumps is considered to be a lower bound for the onset of period-doubling bifurcation. By linearizing the libration orbit around the fixed point, an approximate closed-form bound for the period-doubling route to chaos and the associated minimum of that bound can then be expressed as (see supplementary information for detailed derivation):\\

\begin{equation}
\begin{split}
 {F_2} =& \sqrt{4(\omega_{\mathrm{L}}-\Omega)^2+\gamma^2}\times\lvert A_2-A_{1,3}\rvert\\
 {F_{\mathrm{2min}}} =& \gamma\times\lvert A_2-A_{1,3}\rvert\\
 \end{split}
\label{eqn:Eq3}
\end{equation}
\indent\indent\ where $\mathrm{F_{{2min}}}$ is the minimum required ${F_2}$ necessary to induce period doubling, $\mathrm{\omega_L}$ is the libration frequency, and $A_1$, $A_2$, and $A_3$, are the amplitudes of the three steady-state fixed points, with $A_2$ being the unstable one.\\
\indent\indent\ Eq.~(3) sheds light on the experimental results in Fig.~2: by realising that $\mathrm{\omega_L}<0$ for the low-amplitude branch and $\mathrm{\omega_L}>0$ for the high-amplitude branch, it is easy to understand that period-doubling in Fig.~2(b) takes place mostly for $\Omega\approx\ \mathrm{\omega_L}$.\\
\indent\indent\ A two-dimensional sweep of both $\Omega$ and ${F_2}$ provides further grounds for comparison between the numerical, analytical, and experimental results. Such a sweep is shown in Fig.~3 for both solution amplitude branches and the same drive conditions as in Fig.~2, i.e. $\omega_1/2\pi~=~1566.5$ kHz and $\mathrm{F_1}~=~3~\mathrm{V_{PP}}$. Figure~3 plots the lag of the auto-correlation maximum for the experimentally and numerically obtained time-domain signals, where an auto-correlation lag of 1 indicates P1 orbits, a lag of 2 indicates P2 orbits, and so on. Experimental and numerical data agree well in predicting the region corresponding to P2, the higher order bifurcations, and chaos. It is also interesting to note that both types of time-domain data show that for some parameter-space values the initial condition branch is unstable, and inter-branch jumps occur. Chaos is verified for a selection of experimental traces,
where a correlation dimension \cite{hegger1999practical} of $D_2=2.2\pm0.1$ and $D_2=2.3\pm0.1$ is estimated for the high- and low-amplitudes branches respectively. When a similar analysis is undertaken for simulated time-series, the corresponding largest Lyapunov exponent calculated directly from the differential equations is $\lambda_1=0.141\pm0.002$ and $\lambda_1=0.131\pm0.002$ (see supplementary information).\\
\indent\indent\ Both results presented in Fig.~3 validate the main conclusions of the analytical model; for instance, the P2 bifurcation is mainly obtained with negative detuning for the low amplitude branch and positive detuning for the high amplitude branch, as predicted by the libration frequency analysis. Experimental and numerical results also demonstrate that the model can help constrain the necessary parameter space for chaos generation. It is equally interesting to note that the low-amplitude branch simulations have a well-formed wedge area for period doubling and chaos, and this is nearer to the analytically constrained parameter space compared to the high-amplitude branch. This can plausibly be attributed to the shape of the libration orbits around the low-amplitude branch, which more closely resembles circular ones compared to the almost banana-shaped high-amplitude branch libration orbits.\\

\begin{figure}
	\graphicspath{{Figures/}}
	\includegraphics[width=85mm]{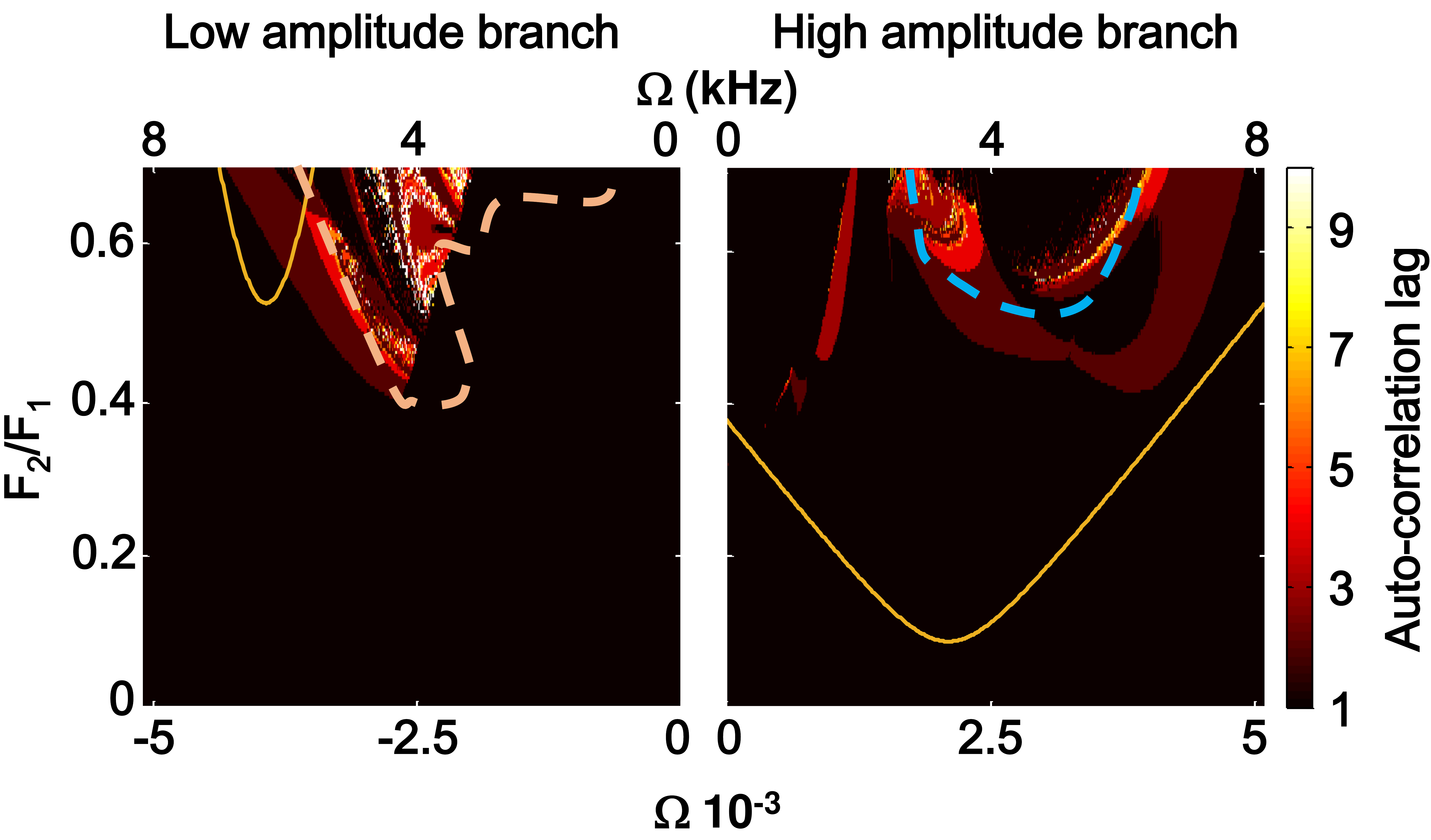}
	\caption{Numerically obtained two-dimensional maps (both panels) showing the location of auto-correlation peak as a function of detuning and forcing ($\Omega$, ${F_2}$). Values greater than 1 (red and bright areas) indicate period-doubling bifurcations and chaos. The experimentally obtained bifurcation areas are equally shown (delineated by the dashed lines). All results are obtained for ${{\delta} = 4.2\times10^{-3}}$ (i.e., $\omega_1/2\pi~=~1566.5$ kHz), and ${{F_1}\sqrt{\alpha} = 1.5\times10^{-4}}$ (i.e., $\mathrm{F_1}~=~3~\mathrm{V_{PP}}$). The area bounded by the analytical model is shown as the solid yellow lines.}
\end{figure}

\indent\indent\ Next, we verify how closely the analytical and numerical results evolve as a function of changing drive conditions (changing $\delta$ or ${F_1}$). Based on the analytical model, it would be expected that, as the edge of the bistable area is approached, the distance between one of the stable fixed points and the unstable fixed point shrinks to zero, and as a consequence, the necessary ${F_2}$ required to achieve P2 and chaos itself is reduced to zero. This is confirmed numerically by performing 3D sweeps ($\delta$, $\Omega$, and ${F_2}$) and tracking the minimum necessary values of ${F_2}$ and $\Omega$ for the onset of P2. These are plotted against $\delta$ (for ${F_1}= 3~\mathrm{V_{PP}}$), along with ${{\mathrm{F_{2min}}}}$ and $\mathrm{\Omega_{min}}$ as obtained from Eq.~(3), in Fig.~4(b)-(c), respectively. Again, numerical results agree with the main features of the analytical model, in that both  ${F_2}$ and $\Omega$ reduce to zero as the saddle node bifurcation is approached. Similarly to the results in Fig.~(3), the analytical results for the low-amplitude branch adhere better to the numerical simulations compared to the high-amplitude branch ones, which show more exotic dynamics, a repeated hint that the eccentric libration orbit is more difficult to model.\\

\begin{figure}
	\graphicspath{{Figures/}}
	\includegraphics[width=85mm]{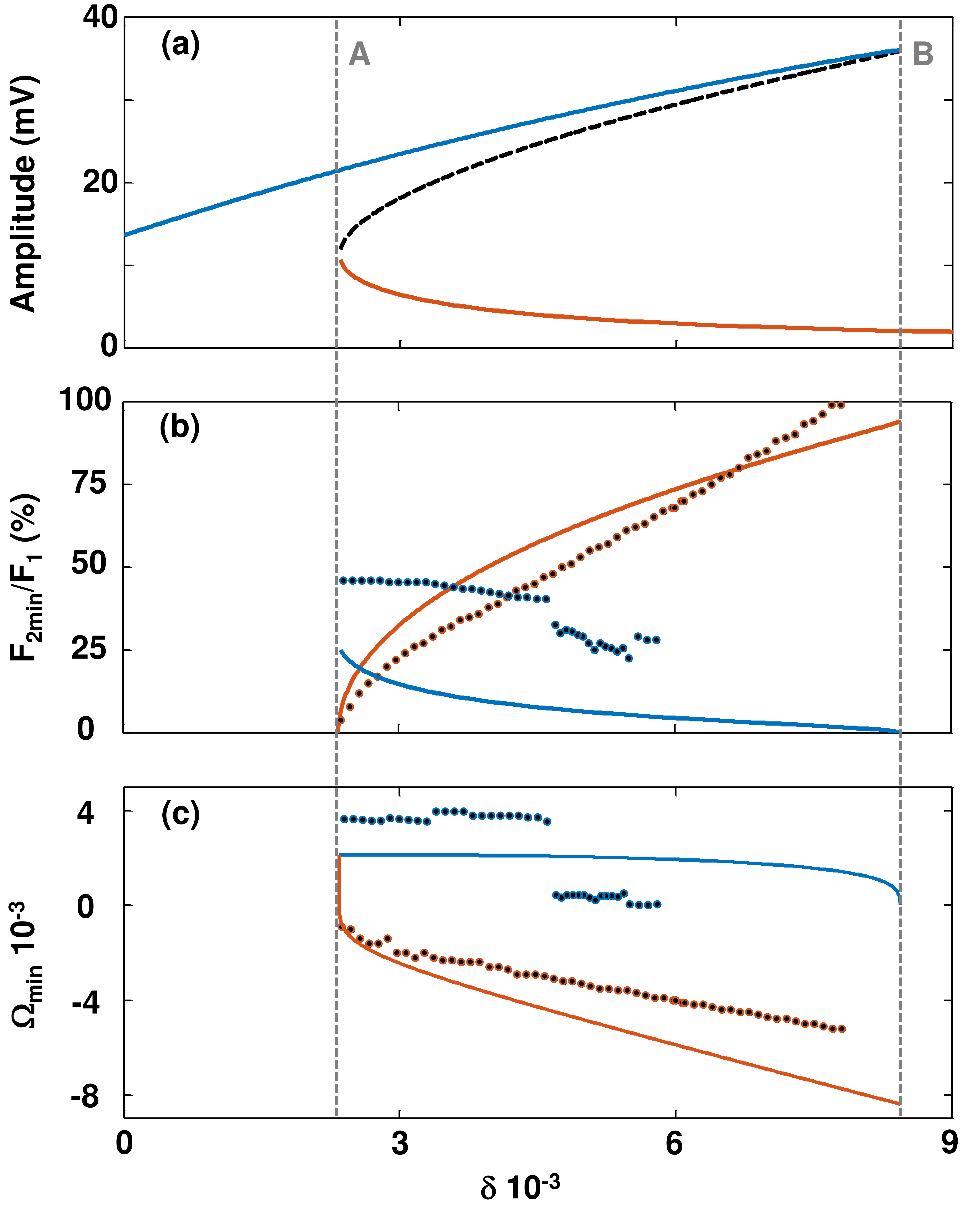}
	\caption{Comparison between numerical (dots) and analytical (solid lines) values of $\mathrm{F_{{2min}}}$ (b) and  $\mathrm{\Omega_{{min}}}$ (c) required for P2 and chaos for both the low amplitude branch (red) and high amplitude branch (blue) of the Duffing resonator shown in (a). For both numerical and analytical data, P2 only appears in the range where bisatbility exists (indicated by the dashed vertical grey lines, and denoted A and B respectively). As the high amplitude branch approaches the saddle-node bifurcation its libration motion becomes more unstable, hence the data points do not reach the saddle-node bifurcation.}
\end{figure}

\indent\indent\ It is interesting to ask whether the perturbed homoclinic-based argument presented above remains valid for the region where bistability is suppressed by the damping (note that the Melnikov approach requires bistability in the lossless version of the system). Indeed, the Melnikov method requires a homoclinic in the undamped system; however, it also assumes that the additional damping and forcing (corresponding to the terms within the brackets in Eq.~2) are small, perturbation-like terms. The authors therefore conjecture that, in the case of suppressed bistability, these terms are sufficiently large to invalidate the perturbed Hamiltonian approach.\\
\indent\indent\ On a practical note, it maybe tempting to pursue chaos generation by positioning the drive near the saddle point bifurcations, as that would require a very small perturbation tone. This, however, is not an optimal experimental condition as it would be easy under the effect of noise to jump to the adjacent potential well and remain stuck there in a low-amplitude libration orbit. It is therefore more favourable for practical ends to position the main tone towards the middle of the bistable region and apply a moderate-amplitude second tone.\\
\indent\indent\ In summary, this work presented an approach to chaos generation in nonlinear M/NEMS resonators that uses drive amplitudes that are typically an order of magnitude smaller than those previously reported. The method relies on applying two drive tones in order to make the rotating-frame phase-space three dimensional, and presents a model that uses linearization of perturbations within the rotating-frame to constrain the parameter space where chaos can be generated. This approach demonstrates that once bistability has been accessed via the first applied tone, chaos can in principle be generated with an arbitrarily small value of a perturbing second tone. The generality of the proposed method, and the relatively low driving forces involved underline applicability to a large range of nonlinear resonators thus potentially placing resonators, particularly M/NEMS ones, as leading candidates for the high-integration physical implementation of networks and reservoirs, as well as the experimental investigation of chaos-related phenomena.\\

\section{\label{sec:ack}Acknowledgements}
 \indent\indent\ This work is partly supported by a MEXT Grant-in-Aid for Scientific Research on Innovative Areas “Science of hybrid quantum systems” (Grant No. JP15H05869 and JP15K21727).\\
\indent\indent\ The work of Ludovico Minati was supported by the World Research Hub Initiative (WRHI), Institute of Innovative Research (IIR), Tokyo Institute of Technology, Tokyo,Japan.\\

\clearpage
\newpage

\title{Supplementary Materials:\\
A Generic Rotating-Frame-Based Approach to Chaos Generation in Nonlinear M/NEMS Resonators}

\author{Samer Houri}
\email{Samer.Houri.dg@hco.ntt.co.jp}
\affiliation{NTT Basic Research Laboratories, NTT Corporation,
3-1 Morinosato-Wakamiya, Atsugi-shi, Kanagawa 243-0198, Japan.}

\author{Motoki Asano}
\author{Hiroshi Yamaguchi}
\affiliation{NTT Basic Research Laboratories, NTT Corporation,
3-1 Morinosato-Wakamiya, Atsugi-shi, Kanagawa 243-0198, Japan.}

\author{Natsue Yoshimura}
\affiliation{Institute of Innovative Research, Tokyo Institute of Technology, Yokohama 226-8503, Japan.}

\author{Yasuharu Koike}
\affiliation{Institute of Innovative Research, Tokyo Institute of Technology, Yokohama 226-8503, Japan.}

\author{Ludovico Minati}
\email{minati.l.aa@m.titech.ac.jp}
\affiliation{Tokyo Tech World Research Hub Initiative (WRHI), Tokyo Institute of Technology, Yokohama 226-8503, Japan.}

\date{\today}

\maketitle

\section{Experimental setup}

The MEMS resonator used is a doubly-clamped beam structure that is 150 $\mu$m$\times$20 $\mu$m$\times$600~nm in size. The material consists of an Al$_{0.3}$G$_{0.7}$As/GaAs heterostructure, the interface of which forms a two-dimensional electron gas (2DEG) layer that is used as one of two transduction electrodes \cite{yamaguchi2017gaas}. The 2DEG layer is contacted using a gold-germanium-nickel diffusive contact. The GaAs layer acts as piezoelectric transducer on top of which a gold layer is deposited to act as a counter-electrode. The device-under-test is wire-bonded and placed in a vacuum chamber (pressure $\sim 10^{-4}$ Pa) with electrical and optical access.\par
Electrical signals are supplied by two synchronized waveform generators (type WF1974, NF Corporation), whereas mechanical motion is detected via a Laser Doppler Vibrometer (LDV, Neoark Corporation) whose output signal is simultaneously fed to a vector signal analyzer (VSA, type HP89410A, Keysight) and a lock-in amplifier (type SR844, SRS). The output of the lock-in amplifier is sampled using a digital oscilloscope (type MSO4034B, Tektronix). A schematic representation of the device and measurement setup is shown in Fig.~(5). The data for the 2D-sweeps shown in the main text are captured using the VSA, since additional oscilloscope data would be prohibitively large. However, some high-resolution long-duration sweeps are experimentally obtained using the oscilloscope for detailed time-series analysis as shown in section III.B.\par
To produce the sweeps shown in Figs.~1-2 of the main text, both applied driving tones are reset before each data point is sampled, to ensure that the system is in the desired branch in case an accidental branch jump took place. This protocol is equally followed in the numerical simulations for the same reasons.\par
\begin{figure}[b]
\graphicspath{{Figures/}}
\includegraphics[width=85mm]{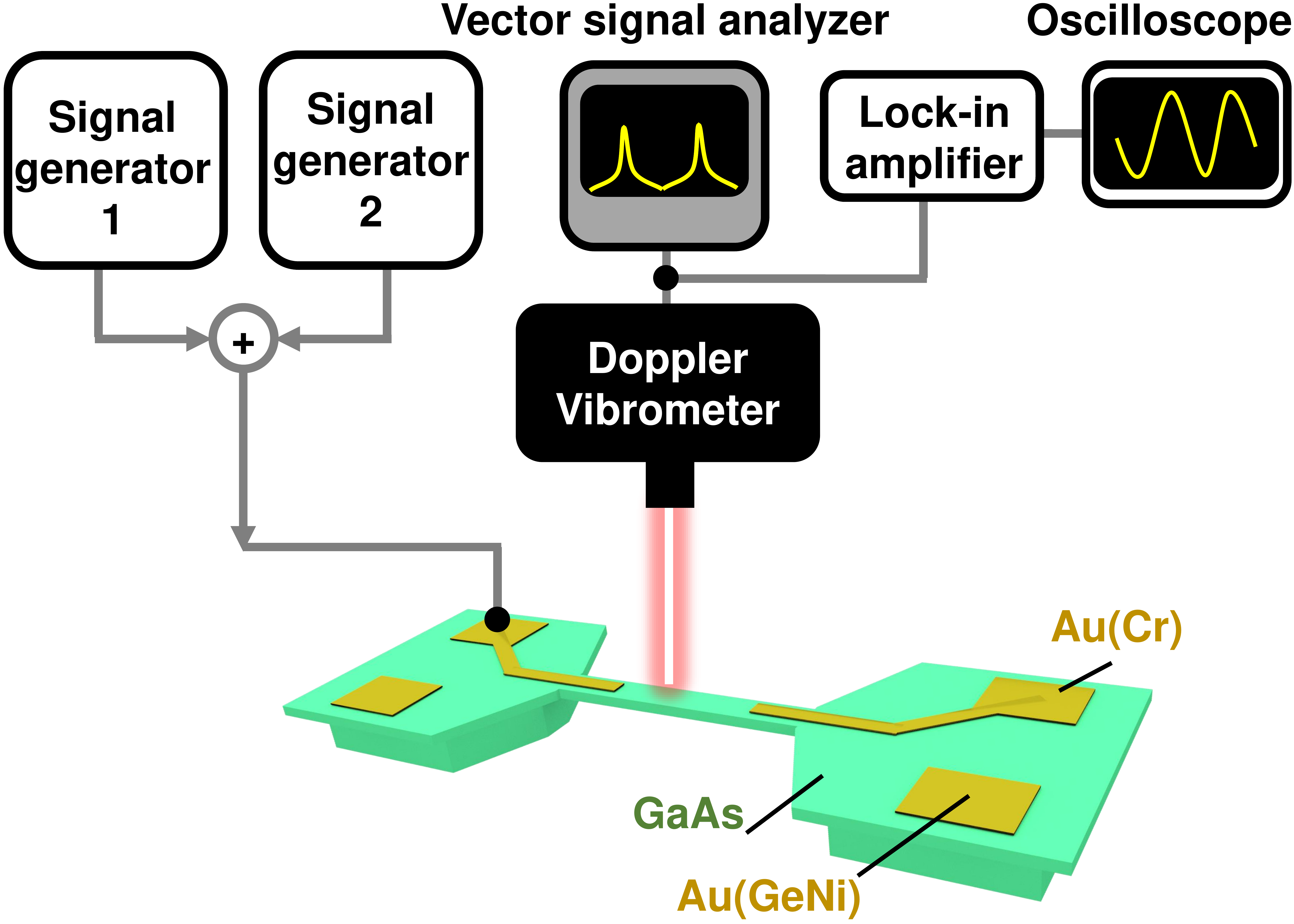}
\caption{Schematic representation of the experimental setup showing the device-under-test, LDV, and other instrumentation. The LDV output is fed to the VSA and lock-in amplifier simultaneously.}
\end{figure}

\begin{figure}[bh]
\graphicspath{{Figures/}}
\includegraphics[width=85mm]{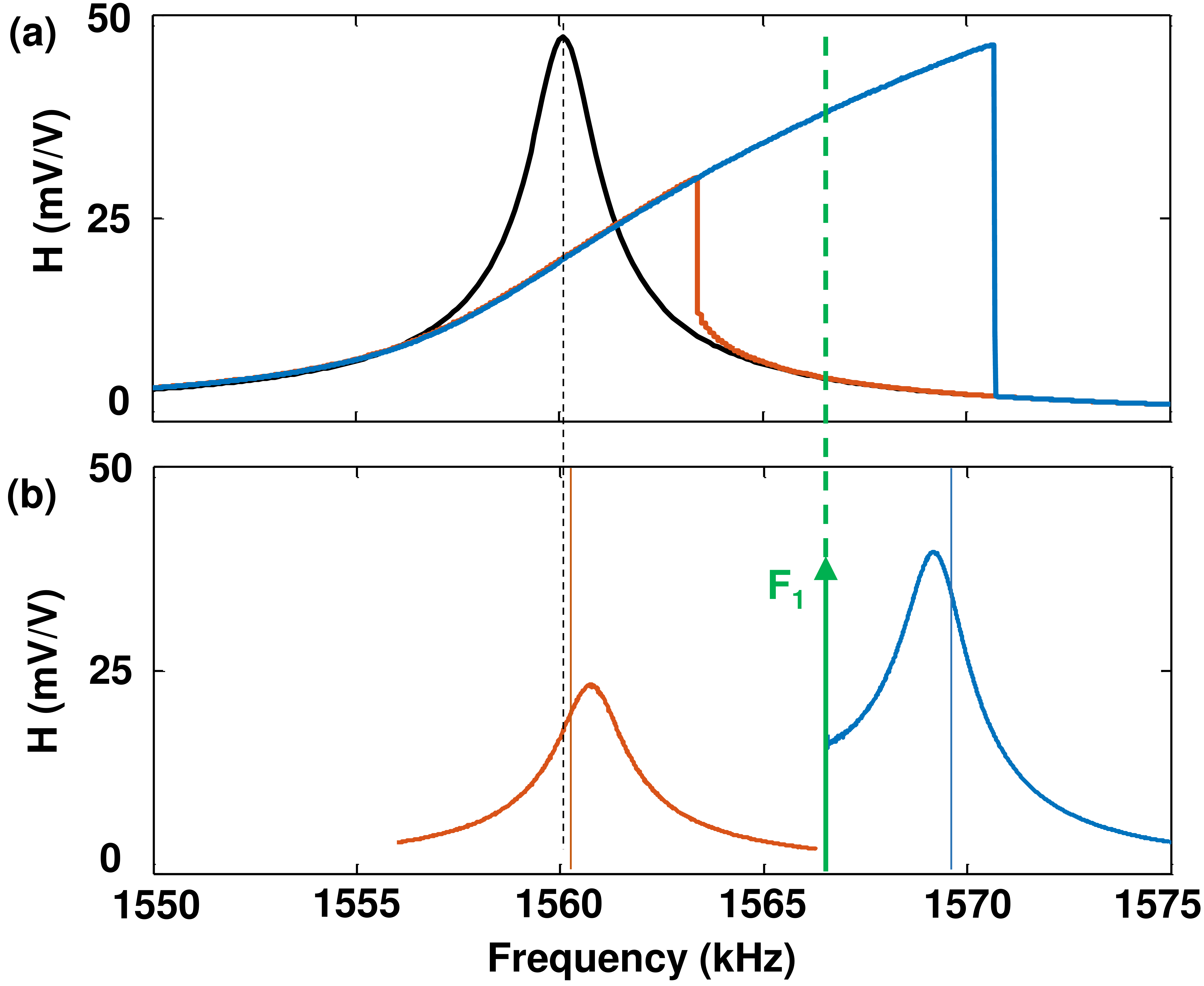}
\caption{Experimentally-obtained libration frequency for the low-amplitude (red) and high-amplitude (blue) branches, measured using a $50~\mathrm{mV_{PP}}$ frequency-swept perturbing tone overlaid on top of an $F_1 = 3~\mathrm{V_{PP}}$ fixed tone denoted by the green line. $H$ denotes the relative amplitude response, expressed in mV per V in drive. The solid vertical lines in panel (b) indicate the location of the libration frequency as obtained from Eq.~(7) for the low-amplitude (red line) and high-amplitude (blue line) branches}
\end{figure}
\section{Derivation of the perturbation model}
In practice, Duffing resonators exist in diversified forms including the following examples: a) having a negative linear component and a positive nonlinear component which is typically used to describe buckled beams \cite{moon1979magnetoelastic,holmes1979nonlinear}; b) having a zero linear component and a positive nonlinear component, such as the system described by \cite{ueda2000strange}; c) having a positive linear component alongside either a positive or a negative nonlinear components (Eq.~(1). In the main text) \cite{cleland2013foundations,greywall1994evading,kozinsky2006tuning,lifshitz2008nonlinear,younis2003reduced,nayfeh1980nonlinear}. The latter encompass archetypal nonlinear M/NEMS resonators \cite{cleland2013foundations,greywall1994evading,kozinsky2006tuning,lifshitz2008nonlinear,younis2003reduced,nayfeh1980nonlinear}, insofar as the displacement is moderate compared to some scaling parameter, e.g. thickness \cite{kaajakari2004nonlinear}. That is indeed the case in the present study and in contrast with the existing works eliciting chaos through large-amplitude forcing \cite{moon1979magnetoelastic}. Remarkably, the Duffing equation is also descriptive of other types of nonlinear resonators such as optical \cite{hansson2014bichromatically}, and superconducting \cite{manucharyan2007microwave} systems.\\
Thus, we consider the Duffing equation representative of nonlinear M/NEMS resonators (Eq.~(1) in the main text) with only one forcing term
\begin{equation}
\ddot{x} + \gamma\dot{x} + \omega_{0}^2x + \alpha x^3 = {F}\cos(\omega_1t)\textrm{ ,}
\label{eqn:EqS1}
\end{equation}
where all quantities take the meaning defined in the main text. We also define $\delta = (\omega_1-\omega_0)/\omega_0$.\\
Because the vibration amplitudes are on the order of the beam thickness, the response of the structure can be analyzed using a perturbation-based technique \cite{cleland2013foundations,kozinsky2006tuning}, whereby the response is assumed to have a sinusoidal component whose amplitude is slowly modulated by a complex envelope. We apply the rotating frame approximation (RFA) following examples set in \cite{cleland2013foundations,nayfeh1980nonlinear}, whereby Eq. (4) is rewritten in the form
\begin{equation}
{\ddot{x} + \epsilon\gamma\dot{x} + \omega_{0}^2x + \epsilon\alpha x^3 = \epsilon{F}\cos(\omega_1t)}\textrm{ ,}
\label{eqn:EqS2}
\end{equation}
where $\epsilon$ is meant to indicate perturbation-order terms.
$x(t)$ is assumed to take the following form\\
\begin{equation}
x(t) = \frac{1}{2}(Ae^{i\omega_1 t}+A^*e^{-i\omega_1 t})\textrm{ ,}
\label{eqn:EqS3}
\end{equation}
Placing Eq.(6) in Eq.(5) and keeping terms only on the order of $\epsilon$, knowing that $\dot{A}/\omega A \approx\mathcal{O}(\epsilon)$, gives the following approximations\\
\begin{equation}
\begin{split}
\ddot{x}~\approx&~\frac{-\omega_1^2}{2}Ae^{i\omega_1 t}+\frac{i\omega}{2} \dot{A} e^{i\omega_1 t}\textrm{,}\\
\gamma\dot{x}~\approx&~\frac{i\gamma\omega}{2}{A} e^{i\omega_1 t}\textrm{,}\\
\alpha x^3~\approx&~\frac{3\alpha}{8}{AA^*A} e^{i\omega_1 t}
 \textrm{,}\\
 \omega_0^2x~\approx&~\frac{\omega_0^2}{2}Ae^{i\omega_1 t} \textrm{,}\\
 \omega_1^2~\approx&~\omega_0^2(1+2\delta) \textrm{.}\\
\end{split}
\end{equation}
Combining the above terms in Eq.(4) gives the following equation in the rotating frame
\begin{equation}
-\delta A+i\dot{A} + \frac{i\gamma}{2}A+\frac{3\alpha}{8}AA^*A=\frac{F}{2}\textrm{ .}
\label{eqn:EqS4}
\end{equation}
Its steady-state solution is obtained by solving
\begin{equation}
-\delta A_0+\frac{i\gamma}{2}A_0+\frac{3\alpha}{8}\lvert A_0\rvert^2A_0=\frac{F}{2}\textrm{ .}
\label{eqn:EqS5}
\end{equation}
Eq.~(9) admits bistability, as shown in Fig.~(1) of the main text for the lossless case (i.e. $\gamma=0$) and a low-loss case ($\gamma=10^{-3}$). The associated rotating-frame pseudo-Hamiltonian corresponding to the steady-state solution of Eq.~(9) is given as in Refs. \cite{dykman2019resonantly,huber2019detecting}
\begin{equation}
\mathcal{H} = -\frac{\delta}{2}(X^2+Y^2) -\frac{F}{2}X  +\frac{3\alpha}{32}(X^2+Y^2)^2\textrm{ ,}
\label{eqn:EqS6}
\end{equation}
where $X$ and $Y$ are the in-phase and quadrature components, respectively, and the complex envelope is written as $A = X + iY$.\par
Furthermore, we assume that $A = A_0 + \bar{A}$, where $\bar{A}$ is a small perturbation term, and $A_0$ is the steady-state solution, which is written as $A_0 = \{A_1,A_3\}$ with the subscripts 1 and 3 indicating the high- and low-amplitude solution branches, respectively. Thus expanding Eq.~(8), bearing in mind that $\dot{A}_0 = 0$, gives
\begin{equation}
\begin{split}
-\delta \bar{A}+i\dot{\bar{A}}+\frac{i\gamma}{2}\bar{A}+\frac{3\alpha}{8}\big(2\lvert A_0\rvert^2\bar{A}+&\\
A_0^2\bar{A}^*+2\lvert A_0\rvert^2A_0+A_0^*\bar{A}^2+ 2\lvert \bar{A}\rvert^2\bar{A} \big)&=0\textrm{ .}
\label{eqn:EqS7}
\end{split}
\end{equation}
We linearize the above equation and drop the damping term, which simplifies the analysis but does not alter the results in a fundamental way, giving
\begin{equation}
    i\frac{\mathrm{d}}{\mathrm{d}t}\begin{pmatrix}
\bar{A}\\
\bar{A}^* \end{pmatrix} = \begin{pmatrix}
\delta-\frac{3\alpha}{4}A_0^2 & -\frac{3\alpha}{8}A_0^2\\
\frac{3\alpha}{8}A_0^2 & -\delta+\frac{3\alpha}{4}A_0^2
\end{pmatrix}
\begin{pmatrix}
\bar{A}\\
\bar{A}^* \end{pmatrix}\textrm{ .}
\label{eqn:EqS8}
\end{equation}
The eigenvalues of the above system, i.e. its libration frequency, are now obtained as
\begin{equation}
\omega_\textrm{L} = \pm\sqrt{\left(-\delta+\frac{3\alpha}{8}A_0^2\right)\left( -\delta+\frac{9\alpha}{8}A_0^2\right)}\textrm{ .}
\label{eqn:EqS9}
\end{equation}
Note that for the high-amplitude branch the libration frequency is $\omega_\textrm{L}>0$, whereas for the low-amplitude branch the libration frequency is $\omega_\textrm{L}<0$, where $\omega_\textrm{L}$ represents the detuning with respect to $\delta$. Experimentally-obtained libration measurements are shown in Fig.~(6).\par
The libration frequency thus represents the natural frequency response of the system to an infinitesimally small perturbation force $F_2$. The onset of the period-doubling bifurcation is considered to be the onset of inter-well jumps. This criterion is approximated by considering the threshold to be the libration amplitude $A_\textrm{L}$ equal to the amplitude difference between the saddle-node and whichever fixed point the system happens to be in, thus $A_\textrm{L}= \lvert A_2 - A_{1,3}\rvert$, which, if we continue to treat the libration motion as linear orbits, gives the equation for the underdamped driven harmonic resonator
\begin{equation}
 {F_2} = \sqrt{4(\omega_\textrm{L}-\Omega)^2+\gamma^2}\times\lvert A_2-A_{1,3}\rvert\textrm{ .}
 \label{eqn:EqS10}
\end{equation}
\section{Melnikov integral}
The purpose of applying the Melnikov method is to obtain a concise and tractable analytical relation.
This is done by first obtaining an analytical formulation for the Homoclinic orbit, then inserting it into the Melnikov integral \cite{holmes1979nonlinear,moon1979magnetoelastic}.\par
Here, we attempt to obtain the analytic formulation in the rotating frame, which can be done by first re-expressing Eq.~(8), after dropping the dissipation term, as
\begin{equation}
\begin{split}
i\frac{\mathrm{d}A}{\mathrm{d}t} =& f(A),\\
f(A) =& \delta A_0-\frac{3\alpha}{8}\big(2\lvert A_0\rvert^2\bar{A}+\\
&A_0^2\bar{A}^*+2\lvert A_0\rvert^2A_0+A_0^*\bar{A}^2+ 2\lvert \bar{A}\rvert^2\bar{A} \big)\textrm{ .}
\label{eqn:EqS11}
\end{split}
\end{equation}
By rearranging and integrating, we obtain
\begin{equation}
i\int_{A(t_0)}^{A(t)}\frac{1}{f(A)}\mathrm{d}A = t-t_0\textrm{ .}\\
 \label{eqn:EqS12}
\end{equation}
Although Eq.~(16) can have a closed-form solution, it is a complicated one, and does not lend itself well to be used in calculating the Melnikov distance.

\begin{figure}
\graphicspath{{Figures/}}
\includegraphics[width=85mm]{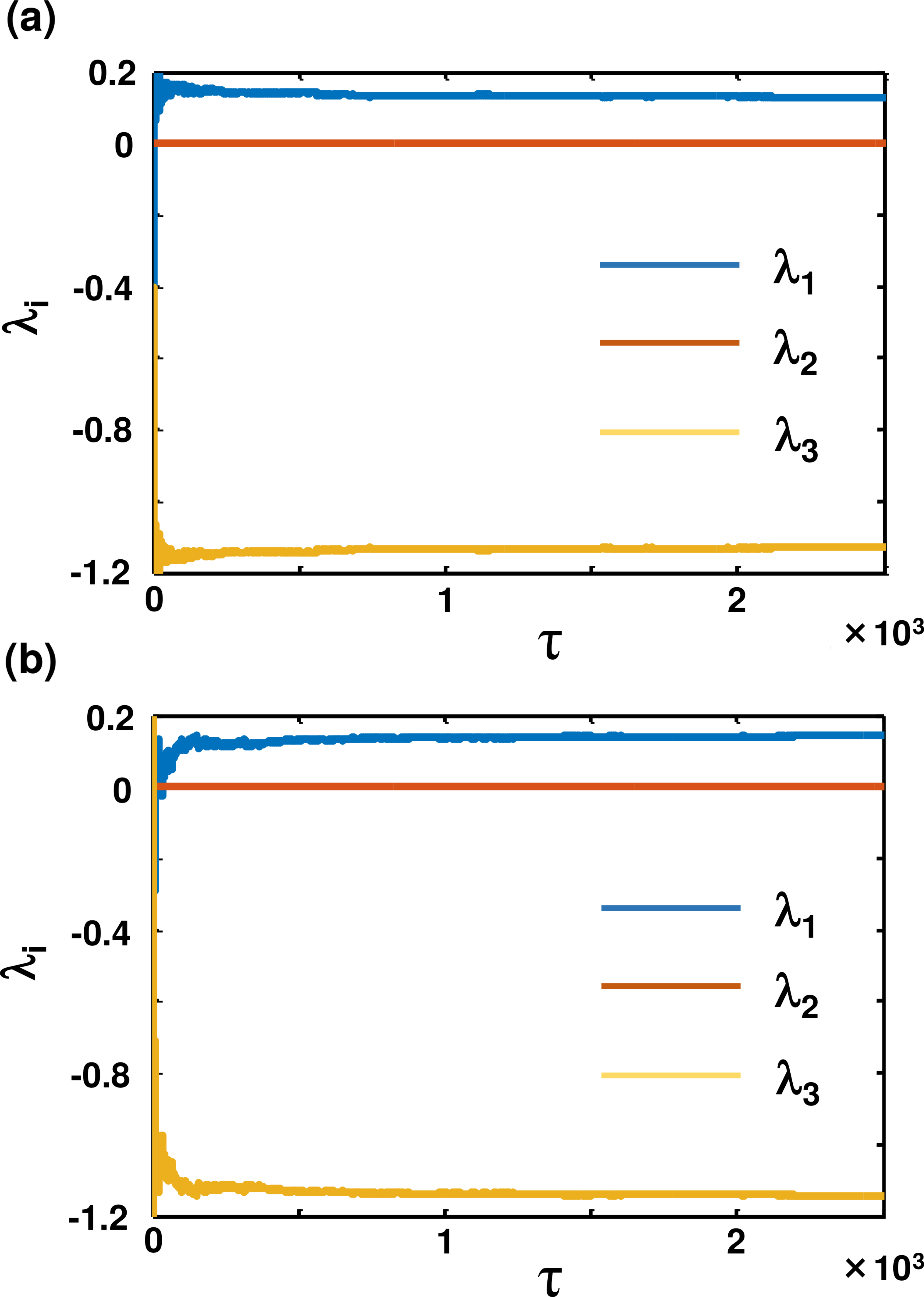}
\caption{Dynamics of the Lyapunov exponents, confirming chaotic behavior in the low-amplitude (a, $\Omega=-0.0031$) and high-amplitude (b, $\Omega=0.004$) branches.}
\end{figure}

\section{Nonlinear time-series analysis}
\subsection{Simulations}
To confirm chaoticity, the spectrum of Lyapunov exponents $\lambda_i$ is calculated directly from the differential equations using an established method based on solving the corresponding variational equation while iteratively applying the Gram-Schmidt orthonormalization \cite{wolf1985}. The Jacobian of the system is
\begin{equation*}
{\bf J} = 
\begin{pmatrix}
-\gamma/2-2\alpha XY & \delta-2\alpha Y^2-\alpha A^2 & F_2/2\cos(\Theta) \\
2\alpha X^2-\delta+\alpha A^2 & 2\alpha XY-\gamma/2 & F_2/2\sin(\Theta)\\
0 & 0 & 0       
\end{pmatrix}\textrm{ ,}
\end{equation*}
where $A^2=X^2+Y^2$, from which the presence of a zero exponent is evident. For convenience of comparison to numerical studies wherein the parameters of the Duffing equation are oftentimes set on the order of unity (e.g., Ref. \cite{sharma2012effects}), here we rescale the physical parameters so that for the low-loss case $\hat{\gamma}=1$, with $\hat{\delta}=10^3\delta$, $\hat{\alpha}=10^3\alpha$, $\hat{F_1}=10^3F_1$, $\hat{F_2}=10^3F_2$, $\hat{\Omega}=10^3\Omega$ and $\tau=10^{-3}t$, where $t$ denotes the normalized time, i.e., $\omega_0=2\pi$. Below, this yields $\hat{\delta}\approx1$ and $\hat{\Omega}\approx1$. The explicit Runge-Kutta (4,5) formula is used, setting relative and absolute tolerance to $10^{-8}$ and $10^{-10}$ respectively.\par
Simulations representative of the physical device are undertaken with $\hat{\delta}=4.2$, $\hat{\alpha}=6513.8$ and $\hat{F_1}=0.036$. For the low-amplitude branch, we set $\hat{F_2}=0.54\hat{F_1}$ and consider a chaotic case having $\hat{\Omega}=-3.1$, alongside a non-chaotic case (period-4) having $\hat{\Omega}=-3.2$. For the high-amplitude branch, we set $\hat{F_2}=0.71\hat{F_1}$ and consider a chaotic case having $\hat{\Omega}=4$, alongside a non-chaotic case (period-6) having $\hat{\Omega}=4.1$. These settings are not critical. All simulations are run until $\tau_\textrm{end}=2.5\times10^3$, performing the orthonormalization in steps of $\tau_\textrm{step}=3\times10^{-3}$; furthermore, to ensure stabilization of the initial transient, all data from $\tau<\tau_\textrm{end}/2$ are discarded.\par
In the low-amplitude branch, chaotic dynamics ($\Omega=-0.0031$) are characterized by rapid convergence to one positive Lyapunov exponent, as visible in Fig. 7(a), with $\lambda_1=0.131\pm0.002$ (mean$\pm$standard deviation), $\lambda_2=0$ and $\lambda_3=-1.131\pm0.002$, yielding a Kaplan-Yorke dimension $D_\textrm{KY}=2.116\pm0.001$. For the non-chaotic case ($\Omega=-0.0032$) we obtain $\lambda_1=0$, $\lambda_2=-0.482\pm0.004$ and $\lambda_3=-0.518\pm0.004$.\par
In the high-amplitude branch, chaotic dynamics ($\Omega=0.004$) are also characterized by rapid convergence to one positive Lyapunov exponent, as visible in Fig. 7(b), with $\lambda_1=0.141\pm0.002$, $\lambda_2=0$ and $\lambda_3=-1.141\pm0.002$, yielding a Kaplan-Yorke dimension $D_\textrm{KY}=2.124\pm0.001$. For the non-chaotic case ($\Omega=0.0041$), we obtain $\lambda_1=0$, $\lambda_2=-0.029\pm0.001$ and $\lambda_3=-0.971\pm0.001$.\par
For the chaotic parameter settings, the time-series of the state variable $X$ in the low and high branches are visible in Fig.~8(a) and 8(d), respectively. Irregular cycle amplitude fluctuations are well-evident, with a marked asymmetry in the high branch. In the low branch, the dynamic appear stationary, whereas in the high branch, an alternation of smaller-amplitude rapid cycles and larger-amplitude slower cycles could be appreciated. The corresponding frequency spectra, visible in Fig.~8(b) and 8(e), are also indicative of chaoticity in that a broad continuous component could be observed at low frequencies, upon which distinct peaks corresponding to the forcing and its harmonics are overlaid. The stroboscopic Poincar\'e sections, taken at rate $\Omega$, visible in Fig.~8(c) and 8(f), are characterized by a simple arc-like topology suggestive of low-dimensional dynamics. The dynamics at the non-chaotic parametric settings are markedly qualitatively different, in that the frequency spectra have a characteristic comb-like appearance, visible for comparison in Fig.~8(b) and 8(e), and the Poincar\'e sections collapse to finite point sets, visible in Fig.~8(c) and 8(f).\

\begin{figure*}
\centering
\graphicspath{{Figures/}}
\includegraphics[width=\textwidth]{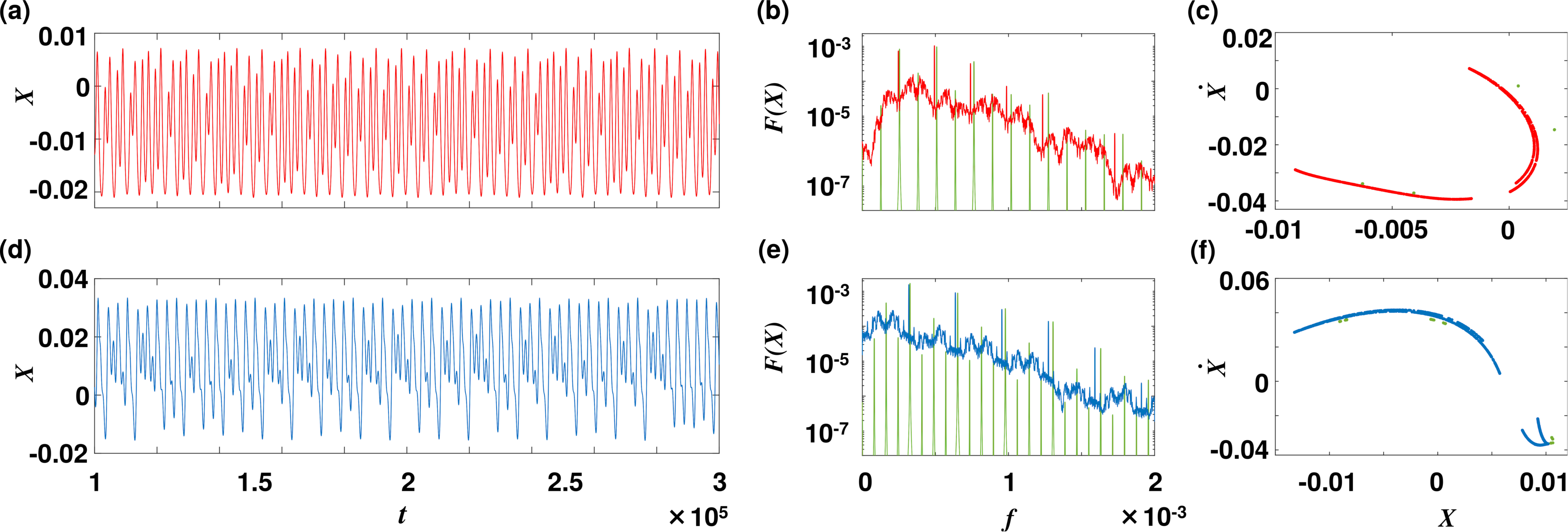}
\caption{Representative simulated time-series for the low-amplitude (a) and high-amplitude (d) branches, alongside corresponding frequency spectra (b, e) and stroboscopic Poincar\'e sections (c, f). Chaotic behavior is observed with $\Omega=-0.0031$ in the low-amplitude branch (red), and with $\Omega=0.004$ in the high-amplitude branch (blue); for comparison, the spectra and sections of two non-chaotic cases (green) are also shown, with $\Omega=-0.0032$ and $\Omega=0.0041$ respectively.}
\end{figure*}

\begin{figure*}
\centering
\graphicspath{{Figures/}}
\includegraphics[width=\textwidth]{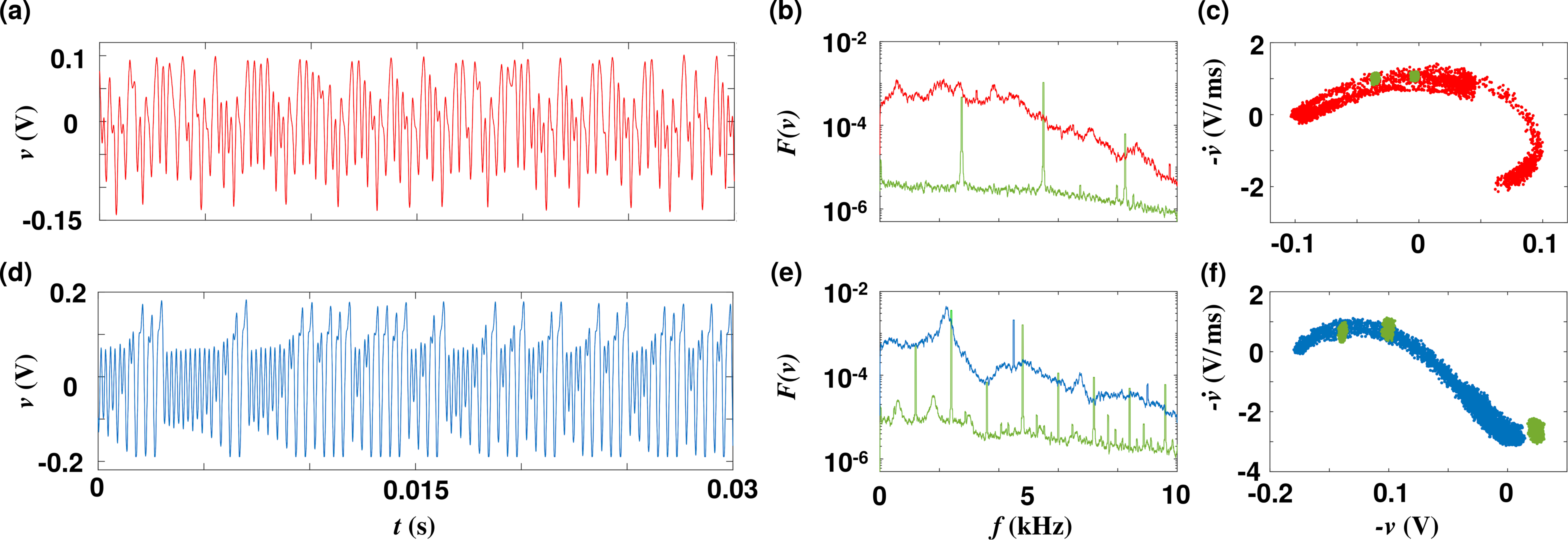}
\caption{Representative experimental time-series for the low-amplitude (a) and high-amplitude (d) branches, alongside corresponding frequency spectra (b, e) and stroboscopic Poincar\'e sections (c, f). Chaotic behavior is observed with $(\omega_2-\omega_1)/2\pi=-3.25\textrm{ kHz}$ ($\Omega=-0.0021$) in the low-amplitude branch (red), and with $(\omega_2-\omega_1)/2\pi=4.5\textrm{ kHz}$ ($\Omega=0.0029$) in the high-amplitude branch (blue); for comparison, the spectra and sections of two non-chaotic cases (green) are also shown, with $(\omega_2-\omega_1)/2\pi=-5.5\textrm{ kHz}$ ($\Omega=-0.0035$) and $(\omega_2-\omega_1)/2\pi=4.8\textrm{ kHz}$ ($\Omega=0.0031$) respectively. Sign inversion applied in (c, f) to ease visual comparison with Fig. 8.}
\end{figure*}

\subsection{Experiments}
Oscilloscope recordings of the in-phase component, digitized at 10 MSa/s, are analyzed using canonical methods based on time-delay embedding as implemented in the \verb+TISEAN+ package (ver. 3.0.1) \cite{hegger1999practical}. The embedding lag $\delta t$ is set to the first local minimum of the mutual information  \cite{fraser1986independent}, and the embedding dimension $m$ is determined via the false nearest-neighbors method ($<5\%$ threshold) \cite{sauer1991embedology}. To reduce the impact of autocorrelation, a Theiler window of width $w$ is set according to the first maximum on the space-time separation plot \cite{provenzale1992distinguishing}. The correlation dimension $D_2$ is thereafter estimated through the Grassberger-Procaccia method, with over-embedding up to a dimension $2m$ and performing a direct search for a convergence plateau \cite{procacia1983measuring,minati2014experimental}. The calculation is repeated over 10 segments of 100,000 points, and the uncertainty determined thereon.\par
An independent confirmation of the attained dynamical complexity is provided via the permutation entropy, a robust rank-based measure; given the map-like series of $l$ local maxima, this is estimated for an sequence order $n=7$, representing the highest value meeting the coverage criterion $l>5n!$, then normalized to $h\in[0,1]$. For this purpose, all available 10 recordings of 10,000,000 points each are pooled to accumulate an adequate number of maxima; no uncertainty is therefore calculated \cite{bandt2002permutation,riedl2013practical}.\par
Accounting for experimental device drift, the following settings are chosen for the low-amplitude branch $\omega_1/2\pi=1566.5$ kHz, $F_1=3$ V and $F_2= 2.1$ V; to obtain chaotic dynamics, we set $\omega_2/2\pi= 1563.25$ kHz, whereas for non-chaotic dynamics (period-2), we set $\omega_2/2\pi= 1561$ kHz. For the high-amplitude branch, the same values of $\omega_1$, $F_1$ and $F_2$ are kept; to obtain chaotic dynamics, we set $\omega_2/2\pi=1571$ kHz, whereas for non-chaotic dynamics (period-3), we set $\omega_2/2\pi= 1571.3$ kHz. All measurements are taken while applying a -1.5 V bias.\par
In the low-amplitude branch, chaotic dynamics ($\Omega=-0.0021$) are associated to a correlation dimension $D_2=2.3\pm0.1$ alongside a permutation entropy $h_7=0.81$; non-chaotic dynamics ($\Omega=-0.0035$) correspondingly feature $D_2=1.8\pm0.2$ and $h_7=0.44$. In the high-amplitude branch, chaotic dynamics ($\Omega=0.0029$) are associated to a correlation dimension $D_2=2.2\pm0.1$ alongside a permutation entropy $h_7=0.62$; non-chaotic dynamics ($\Omega=0.0031$) correspondingly feature $D_2=1.6\pm0.3$ and $h_7=0.44$. Close agreement with simulations in $D_\textrm{KY}\approx D_2$ for the chaotic cases, and the marked difference in permutation entropy between the chaotic and non-chaotic settings, are noteworthy confirmations of the observation of low-dimensional chaos.\par
For the chosen settings yielding chaotic dynamics in the low and high branches, representative time-series are visible in Fig.~9(a) and 9(d). Overall, their qualitative features resemble the simulations remarkably closely, albeit with greater irregularity. The alternation of large- and small-amplitude cycles is well-evident in the high branch. The corresponding frequency spectra, shown in Fig.~9(b) and 9(e), also recall the simulations: for the low branch, a broad component is evident mainly below $f<5\textrm{ kHz}$, whereas for the high branch two peaks are also clearly identifiable, namely, a broad peak at $f\approx2.25\textrm{ kHz}$ and a narrow line at $f\approx4.5\textrm{ kHz}$, corresponding to $\Omega$. The Poincar\'e sections, visible in Fig.~9(c) and 9(f), also delineate an arc-like topology similar to the simulations. Again, for the non-chaotic settings there are marked qualitative differences, in that the frequency spectra have a characteristic comb-like appearance, visible for comparison in Fig.~9(b) and 9(e), and the Poincar\'e sections collapse to finite point sets, visible in Fig.~9(c) and 9(f). Altogether, these results affirm the chaotic nature of the experimental system unquestionably.

\bibliographystyle{apsrev4-1}
\bibliography{apssamp}

\end{document}